\documentclass[10pt,
showpacs,preprintnumbers,amsmath,amssymb,
aps,prd,nofootinbib,eqsecnum,a4paper,
superscriptaddress]
{revtex4}
\pdfoutput=1

\usepackage{graphicx,epsf}
\usepackage{subfigure}
\usepackage{booktabs}

\def\be{\begin{equation}}
\def\ee{\end{equation}}
\def\bea{\begin{eqnarray}}
\def\eea{\end{eqnarray}}
\def\p{\partial}

\def\R{{{\cal{R}}}}
\def\P{{{\cal{P}}}}

\def\vp{{\varphi}}
\newcommand\dvp[1]{{\delta\varphi_{#1}}}
\def\H{{\cal H}}
\def\cs2{c_{\rm{s}}^2}
\def\dPnad{\delta P_{\rm{nad}}}
\def\U0{{\bar U_0}}

\def\N{{\cal{N}}}

\def\vpb{\varphi_0}

\def\bi{\begin{itemize}}
\def\ei{\end{itemize}}
\newcommand\eq[1]{Eq.~(\ref{#1})}

\def\fl{{\rm{flat}}}

\begin{document}

\title{Quantifying the behaviour of curvature perturbations during inflation}

\author{Ellie Nalson}
\email[]{e.nalson@qmul.ac.uk}
\affiliation{
Astronomy Unit, School of Physics and Astronomy, Queen Mary University
of London, Mile End Road, London, E1 4NS, UK}
\author{Adam J.~Christopherson}
\affiliation{School of Physics and Astronomy, 
University of Nottingham, University Park,
Nottingham, NG7 2RD, UK}
\author{Ian Huston}
\author{Karim A.~Malik}
\affiliation{
Astronomy Unit, School of Physics and Astronomy, Queen Mary University
of London, Mile End Road, London, E1 4NS, UK}
\date{\today}
\begin{abstract}
How much does the curvature perturbation change after it leaves the
horizon, and when should one evaluate the power spectrum? To answer
these questions we study single field inflation models numerically,
and compare the evolution of different curvature perturbations from
horizon crossing to the end of inflation. In particular we calculate
the number of efolds it takes for the curvature perturbation at a
given wavenumber to settle down to within a given fraction of their
value at the end of inflation.
We find that e.g.~in chaotic inflation, the amplitude of the comoving
and the curvature perturbation on uniform density hypersurfaces differ
by up to 180 \% at horizon crossing assuming the same amplitude at the
end of inflation, and that it takes approximately 3 efolds for the
curvature perturbation to be within 1 \% of its value at the end of inflation.
\end{abstract}

\pacs{98.80.Cq \hfill  arXiv:1111.6940 
}

\maketitle

\section{Introduction}
\label{sec:introduction}

Our understanding of the early universe has rapidly improved over the
last couple of decades. Observations indicate that we live in a
universe that agrees remarkably well with the cosmological standard
model, a universe that is homogeneous and isotropic
on the largest scales, described by the Friedmann-Robertson-Walker
(FRW) spacetime and that includes cold dark matter and a cosmological constant. 
On smaller scales, anisotropies in the Cosmic
Microwave Background (CMB) and the Large Scale Structure (LSS) are
sourced by quantum fluctuations formed in scalar fields during an
inflationary epoch at very early times. As the process describing the decay of
these scalar fields into standard matter is not known, we map the
power spectrum of the field fluctuations onto the spectrum of a
conserved quantity when the size of
the fluctuations is similar to the size of the Hubble radius, that is
at horizon crossing, $k=aH$, (see e.g.~Refs.~\cite{Kolbandturner,LLBook} for reviews
of inflationary cosmology).
The relative size of a given field fluctuation compared to the Hubble
radius ($1/aH$) is important.
Perturbations arising from fluctuations in the inflaton scalar field
will become fixed, or are ``frozen in'', when that scale first leaves
the Hubble radius.
They will be conserved while that mode is beyond the
Hubble radius if mapped to a suitable quantity like a curvature
perturbation, such as those discussed in the following.  

Observational cosmology is entering an era in which the data from
observations of LSS and the CMB are becoming much more detailed. Only
eight years ago, the WMAP team were quoting cosmological parameters to
an accuracy of about 10\% , Ref. \cite{WMAP1}. Now, as data sets are
improving both in quality and size, the WMAP seven year observations
(hereafter WMAP7) can constrain these parameters to within a couple of
percent, Ref. \cite{WMAP7}. With {\sc Planck} we expect to do even
better. Hence it is essential that the quantities we wish to study in
the early universe are understood, theoretically, to this same level
of precision.

It is well known that the curvature perturbations on both uniform density
hypersurfaces, $\zeta$, and on comoving hypersurfaces, $\R$, are
conserved on large scales where gradient terms can be
neglected (i.e. in the limit $k \to 0$) for adiabatic perturbations, Refs.
\cite{Bardeen83,Lyth85}. This result follows from the conservation of energy, Ref. \cite{WMLL}. 
The standard approach used to calculate the power spectrum of perturbations
after horizon crossing assumes that the limit $k\to 0$ has been reached, see Ref. \cite{Copeland:1993jj}.
However, immediately after horizon crossing the wavenumber will not yet have become sufficiently small
for this limit to be accurate and gradient terms will still play a role.
In fact, in single field inflation there will still be some residual non-adiabatic pressure perturbation,
$\delta P_\mathrm{nad}$, present Ref. \cite{Mollerach}. So, even in the absence of other sources of $\delta P_\mathrm{nad}$,
the curvature perturbation will continue to evolve for some number of
efolds before settling down to its value at the end of inflation.
Although it is well known that this evolution continues for a short time after horizon crossing, the exact amount of evolution 
has not been quantified. 
Exactly how long the evolution will last and how big the errors may be if the curvature perturbation is evaluated too early
are issues which are yet to be addressed in the literature. 
\\

Much work has been done calculating the power spectra for the curvature
perturbations, $\zeta$ and $\R$, during
inflation (see for example Refs.~\cite{Stewart:1993,Grivell,Huang:2000bh,Leach:2001zf,Leach:2001,Stewart:2002}   
and the reviews Refs. \cite{J.Lidsey,hep-ph/9807278}).  
Analytic studies have to rely on
either the slow roll limit, or large scale approximations, to make the
calculations viable.  In these limits, and without anisotropic stress,
the two definitions are equal up to a sign difference
e.g.Ref. ~\cite{MW2008}.  In this paper we will quantify the evolution of
the curvature perturbation shortly before, during and after horizon
crossing. In Section~\ref{sec:results} we will demonstrate how $\zeta$
and $\R$ differ at these times, how much the instantaneous horizon
crossing values differ from the values at the end of inflation and how
long it takes for the quantities to reach these final values.

As we do not want to rely on the slow roll approximation or the large
scale limit, we solve the Klein-Gordon equations numerically, having
chosen a particular scalar field potential.  The linear scalar
field perturbations are evolved from an initial Bunch-Davies state
well inside the horizon until the inflationary expansion ends.  Most
of the results we present in this paper use single field chaotic inflation, the simplest single field
inflation model which is in agreement with WMAP7.
In order to ensure that our results are representative beyond this
simplest model, we also study a set of more complicated single field
models, $U = U_0 + \frac{1}{2}m^2\vp^2$, $U = \frac{1}{4}\lambda
\vp^4$ and $U = \sigma \vp^{2/3}$.  We compare these numerical results
with the standard analytic solutions for single field inflation.  It
is important to recognise that the analytic solution is derived using
the $k\to 0$ limit but is evaluated using quantities at horizon
crossing. Although this has been known in the literature for many
years it is not often made clear when this mixing of late time
solution and horizon crossing values is being used. In this
paper we highlight the large magnitude of the inaccuracies which would
result if a na\"ive calculation of the power spectrum at horizon
crossing is performed using horizon crossing values.\\

Throughout this paper, although we use quantum initial conditions, we only consider the evolution of the scalar field perturbations classically. 
There are issues concerning how and when the quantum-to-classical transition takes place but we will not attempt to address these here.
For further discussion on these issues, see for instance, Ref. \cite{Polarski:1995, Lyth:2006}.  \\

The paper is organised as follows. In Section \ref{sec:equations} we
define the relevant variables and present the relevant governing
equations to determine their evolution. In the following section we
detail the numerics we use to solve the equations. In Section
\ref{sec:results} we present our results, and we conclude in the final
section.

\section{Equations}
\label{sec:equations}

In this section we will derive the equations for the curvature
perturbation and $\dPnad$. Throughout the calculations we expand
quantities into background and perturbative contributions, and shall
consider scalar perturbations about a homogeneous, isotropic, FRW
background model. To first order this leads to the line element
\be
\label{metric}
ds^2 = -a^2\left(1+2\phi\right)d\eta^2 + 2 a^2 B_{,i} d\eta dx^i + a^2\left[\left(1-2\psi\right)\delta_{ij} +2E_{,ij}\right]dx^i dx^j\,,
\ee
where $a=a(\eta)$ is the scale factor, $\delta_{ij}$ is the flat
background metric, $\phi$ the lapse function, and $\psi$ the
curvature perturbation, $B_1$ and $E_1$ are scalar perturbations
describing the shear ($\sigma_{s}\equiv -B+E'$), and $\eta$ is
conformal time, related to coordinate time $t$ by $ad\eta=dt$ (for
details on perturbation theory see e.g.~Ref.~\cite{MW2008}, the
notation of which we follow where possible).
Derivatives
with respect to conformal time are denoted by a dash. Greek indices,
$\mu,\nu,\lambda$, run from $0,\ldots 3$, while lower case Latin
indices, $i,j,k$, run from $1,\ldots3$.
We shall employ the flat slicing and threading (or flat gauge)
throughout this paper, where not stated otherwise.

\subsection{Governing equations}
\label{sec:govequ}

In this paper we only consider single field inflation, which is
governed by the Klein-Gordon equation, Ref. \cite{LLBook}, in the background,
\bea
\label{KGback}
\vp_{0}''+2\H\vp_{0}'+a^2 U_{,\vp}=0\,,
\eea
where $\H$ is related to the Hubble
parameter by $\H\equiv aH$, and prime denotes a derivative with respect
to conformal time $\eta$, and the Friedmann equation,
\be
\label{Friedmann}
\H^2=\frac{1}{3M_{\rm{PL}}^2}  
\left(\frac{1}{2}{\vp'_{0}}^2+a^2 U_0\right)\,.
\ee
By perturbing the field and using the metric defined above in
\eq{metric}, we arrive at the perturbed evolution equation in Fourier
space, 
\be
\label{KG1_flat}
\delta\vp''+2\H\delta\vp'+k^2\delta\vp
+a^2\left\{
U_{,\vp\vp}
+\frac{1}{\H M_{\rm{PL}}^2}\left(
2\vp_{0}'U_{,\vp}
+{\vp_{0}'}^2\frac{1}{\H M_{\rm{PL}}^2}U_0
\right)
\right\}\delta\vp=0\,,
\ee
where we have chosen the flat gauge. Note that in flat gauge the
field fluctuation is also known as the Sasaki-Mukhanov variable, Ref.
\cite{Sasaki1986,Mukhanov88}.
The Fourier component of the field fluctuation, $\delta\vp (k^i)$, is
related to the fluctuation in real space, $\delta\vp (x^i)$, by
\be
\delta\vp (\eta,x^i) = \frac{1}{(2\pi)^3} \int d^3 k \delta\vp (k^i) e^{ik_ix^i} \, .
\ee

In order to evaluate the curvature perturbations we will need to
relate the pressure, $P_0$ and $\delta P$, the energy density $\rho_0$
and $\delta\rho_1$ to the scalar field, $\vp_0$ and $\dvp1$, by
\bea
P_0 &=& \frac{1}{2a^2}{\vp_{0}'}^2- U_0\,,\qquad
\rho_0 = \frac{1}{2a^2}{\vp_{0}'}^2+ U_0\,,
\eea
and
\bea
\delta\rho &=& \frac{1}{a^2}\vp_{0}'\delta\vp'
+\left(U_{,\vp}-\frac{3\H}{a^2}\frac{{\vp_{0}'}^3}{{\vp_{0}'}^2+2a^2U}\right)\dvp{}\,, \qquad
\delta P = \frac{1}{a^2}\vp_{0}'\delta\vp'
-\left(U_{,\vp}+\frac{3\H}{a^2}\frac{{\vp_{0}'}^3}{{\vp_{0}'}^2+2a^2U}\right)\dvp{}\,,
\eea
see e.g.~Ref.~\cite{MW2008} for details.

\subsection{Conserved quantities and $\dPnad$}
\label{sec:consqu}

As pointed out above we map the initial scalar field fluctuations,
which themselves would evolve after horizon exit, onto conserved
quantities, that remain constant in the limit $k \to 0$ for adiabatic
perturbations. We focus here on the curvature perturbation on uniform
density hypersurfaces $\zeta$, Ref. \cite{Bardeen83}, and the comoving curvature
perturbation $\R$, Ref. \cite{Lyth85}.
The spectrum of the curvature perturbations can then be used to set the
initial conditions for standard Boltzmann codes see e.g.~Ref.~\cite{camb},
 that are used to calculate the CMB anisotropies.
The non-adiabatic pressure perturbation $\dPnad$ is in general not
directly observable, but is a source term for the evolution of the
curvature perturbations, see Ref. \cite{WMLL}.
Now we use the equations in the previous section above to find
expressions for the curvature perturbations and the
non-adiabatic pressure perturbation $\dPnad$.

The total pressure perturbation is split into an adiabatic and
non-adiabatic part as Ref. \cite{KS}
\be
\dPnad = \delta P - \cs2 \delta\rho\,,
\ee
where $\cs2$ is the adiabatic speed of sound and is defined as $\cs2
\equiv {P_0}'/{\rho_0}'$.  This gives us an expressions for $\dPnad$
in terms of the scalar field quantities, see Ref. \cite{nonad}
\be 
\dPnad=\Bigg[\frac{U_{,\vp}}{3\H^2 M_{\rm{PL}}^2}\vpb'^2
-2U_{,\vp}\Big(1+\frac{U_{,\vp}a^2}{3\H\vpb'}\Big)\Bigg]\dvp{}
-\frac{2U_{,\vp}}{3\H}\dvp{}'\,.
\ee

The curvature perturbation on uniform density hypersurfaces is defined as
\be
\label{defzeta}
-\zeta \equiv \psi+\frac{\H}{\rho_0'}\delta\rho\,,
\ee
which simplifies if we evaluate the right hand side in flat gauge to
%
%
$-\zeta=\frac{\H}{\rho_0'}\delta\rho_{\fl}\,,$
%
where for ease of use we drop the subscript $\fl$ in the
following. The comoving curvature perturbation, that is the curvature
perturbation evaluated on comoving or uniform field slices, is defined
by
\be
\label{defR}
\R \equiv \psi+\frac{\H}{\vp_0'} \delta\vp\,,
\ee
which simplifies again if the RHS is evaluated in flat gauge to
$\R=\frac{\H}{\vp_0'} \delta\vp$.
Note, that these two gauge-invariant curvature perturbations, defined in different gauges, are related by the constraint equation,
\be
\label{constraint}
k^2 \Psi = -9 \frac{\H^2 {\vp_0'}^2}{2a^2\rho_0} (\R + \zeta)\,,
\ee
where $\Psi = \psi + \H\sigma_{s}$ is the curvature perturbation
in longitudinal gauge, that is on uniform shear hypersurfaces. As can
be seen from \eq{constraint}, $\zeta + \R$ will become small on
super-horizon scales. In the work that follows as well as studying
how quickly $\R$ and $\zeta$ settle to their final conserved
values we will also be looking at how quickly after horizon crossing
these two descriptions of the curvature perturbation agree.

Combining our results above we finally get for an arbitrary wavenumber
$k$ for a single scalar field and without imposing slow-roll in terms
of the scalar field variables the non-adiabatic pressure
\be 
\label{deltaPnadfinal}
\dPnad=-\frac{2 U_{,\vp} }{3 \H }
\left[\delta\vp' +
\left(\frac{a^2U_{,\vp}}{\vp_0'}+\frac{6\H Ua^2}{{\vp_{0}'}^2+2Ua^2}
\right){\delta\vp}
\right]\,,
\ee
and similarly the curvature perturbation on uniform density hypersurfaces,  
\be
\zeta=\frac{1}{3{\vp_{0}'}^2}
\left[
\vp_{0}'\delta\vp'
+\left(U_{,\vp}a^2-3\H\frac{{\vp_{0}'}^3}{{\vp_{0}'}^2+2Ua^2}\right)\dvp{}
\right]\,,
\ee
and the comoving curvature perturbation, defined in \eq{defR}, was already given in terms of the field $\vp$. 


\subsection{Analytic Solutions}
\label{sec:analytic}

Using numerical techniques we can evaluate the expressions above at
any time, analytically this is not possible. 
To deduce an analytic expression for the
curvature perturbation, such as that used in the popular $\delta N$ formalism,
\footnote{
The $\delta N$ formalism uses the direct relation of the
curvature perturbation to the perturbed number of efolds on large
scales, see, for instance, Ref. \cite{Starobinsky:1982ee,Starobinsky:1986fx,Sasaki:1995aw,LMS}.
We can relate the linearly perturbed number of expansions to the density by
\be
\delta N = \frac{\p N}{\p\rho} \delta\rho + \frac{1}{2}\frac{\p^2N}{\p\rho^2} \delta\rho^2 + ... \,.
\ee
} 
we start from the definition \cite{LLBook}
\be
\label{powerdef}
\P_{\R}(k) = \frac{k^3}{2 \pi^2} \left|\R\right|^2 \,.
\ee
By obtaining an exact solution for power-law inflation and taking the $k \to 0$ limit, and then making
an expansion about this solution, we arrive at the following, see Ref. \cite{J.Lidsey}
\be
\label{PRest2}
\P_{\R}(k)_{\rm{est}2*} = [1 - (2C + 1)\epsilon_{SR} + C\eta_{\rm SR}]^2 
\frac{a^2 H^4}{(2\pi)^2 {\vp'_0}^2} 
\Big|_{k=aH}\,,
\ee
where $C = -2 + ln2 + \gamma$ and $\gamma$ is the Euler constant and $\epsilon_{SR}$ and $\eta_{SR}$ are slow roll parameters defined by,
\bea
\epsilon_{SR} &\equiv& \frac{3{\vp'_0}^2}{2} \left[ a^2 U + \frac{1}{2} {\vp'_0}^2 \right]^{-1} = 2M_{\rm{PL}}^2 \left( \frac{H_{,\vp}}{H} \right)^2 \, ,\\
\eta_{SR} &\equiv& 1 -\frac{\vp''_0}{aH\vp'_0} = 2M_{\rm{PL}}^2 \frac{H_{,\vp\vp}}{H} \, .
\eea
The expression given in \eq{PRest2} is only valid in the large scale
limit, or equivalently a long time after horizon exit. However, it
must be evaluated exactly when the corresponding mode crosses the
horizon. In order to arrive at an expression which is valid at all times, we would need to consider the
full analytic solution, including log corrections, see for instance Ref. \cite{Burrage:2011}.
We shall return to this subject in Section
\ref{sect:highlight}.

The above result holds, to the lowest order in slow roll, for any general
potential and if the slow roll parameters are assumed to be very small
this simplifies to
\be
\label{PRest1}
\P_{\R}(k)_{\rm{est}1*} = \frac{a^2 H^4}{(2\pi)^2 {\vp'_0}^2} \Big|_{k=aH}\,.
\ee
For further discussions on the analytic treatment of curvature perturbations close to Horizon crossing, 
see for example Ref. \cite{Kinney:2005, Byrnes:2009}

\section{Numerical Setup}
\label{sec:numerical}

To set up the numerical system we have followed the work done by
Salopek et al. Ref. ~\cite{Salopek:1988qh}. We select a finite range of k modes which
will cover all the modes which have been observed in the CMB. The WMAP
team have released results corresponding to the range $k \in
[3.5$x$10^{-4} , 0.12] Mpc^{-1}$ so we will consider a similar range
below. We use the number of efolds, $\N = log(a/a_{\rm{init}})$, as our
time variable instead of conformal time, where $a_{\rm{init}}$ is the value
of $a$ at the start of inflation and is evaluated by setting $a = 1$
today and using the background run, assuming instantaneous reheating.

The initial conditions for the background system are selected
depending on the choice of potential. For the potential $U =
\frac{1}{2} m^2 \vp ^2$ with $m = 6.32$ x $10^{-6}M_{\rm{PL}} $, we use initial
conditions $\vp_0 = 18M_{\rm{PL}}$ and $\vp_{0,\N} = -0.1M_{\rm{PL}}$, see Ref.
\cite{Ianthesis}.  Following Ref.~\cite{Salopek:1988qh} we set the
initial conditions for each k mode a few efolds before horizon
crossing when the initial time $\N_{\rm{init}}(k)$ is such that 
\be
\frac{k}{aH|_{\rm{init}}} = 50 
\ee 
At early times we assume the
Bunch-Davies vacuum and we get the following initial conditions for
modes well inside the horizon, see Ref. \cite{Salopek:1988qh,Huston:2009ac,Ianthesis}:
\begin{eqnarray}
 \delta\varphi |_{\rm{init}} &=& \frac{1}{aM_{\rm{PL}}\sqrt{2k}}e^{-ik\eta} \, ,  \nonumber \\
 \delta{\varphi}_{,\mathcal{N}} |_{\rm{init}} &=& -\frac{1}{aM_{\rm{PL}}\sqrt{2k}}e^{-ik\eta} \left( 1 + i \frac{k}{aH} \right) \, ,
\end{eqnarray}
where $ \eta = -(aH(1-\epsilon_H))^{-1}$ and $\epsilon_H = -\frac{H_{,\mathcal{N}}}{H}$ \, .

The behaviour of all the modes over the scales we consider are
similar, so for clarity only $k_1$, $k_2$ and $k_3$ modes, given
below, have been shown on the graphs. It is worth noting that $k_2$ is
the WMAP pivot scale.

\bea
k_1 &=& 2.77 \mbox{ x } 10^{-5} Mpc^{-1} = 7.28 \mbox{ x } 10^{-62} M_{\rm{PL}} \\
k_2 &=& 2.00 \mbox{ x } 10^{-3} Mpc^{-1} = 5.25 \mbox{ x } 10^{-60} M_{\rm{PL}} \quad (WMAP)\\
k_3 &=& 1.45 \mbox{ x } 10^{-1} Mpc^{-1} = 3.80 \mbox{ x } 10^{-58} M_{\rm{PL}}
\eea

The results presented below are for the potential $U = \frac{1}{2} m^2
\vp ^2$ with $m = 6.32$ x $10^{-6}M_{\rm{PL}} $, this value has been chosen such that 
$\P_{\R}(k) = 2.45$ x $10^{-9}$ at the end of inflation for the WMAP pivot scale.
 We also obtained results for
three additional potentials, $U = U_0 + \frac{1}{2}m^2\vp^2$, $U =
\frac{1}{4}\lambda \vp^4$ and $U = \sigma \vp^{2/3}$, using parameters
and initial conditions specified in Ref. \cite{Huston:2009ac,Ianthesis}. All the numerical results
have been verified with a second numerical program to ensure their
accuracy, see Ref. \cite{pyflation}.

\section{Results}
\label{sec:results}

Having setup the governing equations and the numerical formalism in
the previous sections, we can now turn to answering the questions we
raised in the introduction.  Here we present our results which
quantify the difference between $\zeta$ and $\R$, the magnitude of
error incurred if we would use \eq{powerdef} evaluated at horizon
crossing instead of \eq{PRest2}, and the length of time taken for the
power spectra to settle to their values at the end of inflation.

\begin{figure}[ht!] 
  \begin{center}
    \mbox{
      \subfigure
[\label{fig:1a} The evolution of the power spectrum of $\zeta$, $\P_{\zeta}(k)$ and $\R$, $\P_{\R}(k)$ is plotted against the number of efolds, $\N$.
Both power spectra stop evolving shortly after horizon crossing, however a short period of evolution immediately after horizon crossing is visible as is
a difference between $\P_{\zeta}(k)$ and $\P_{\R}(k)$.
]{\scalebox{0.5}{\includegraphics[width=160mm]{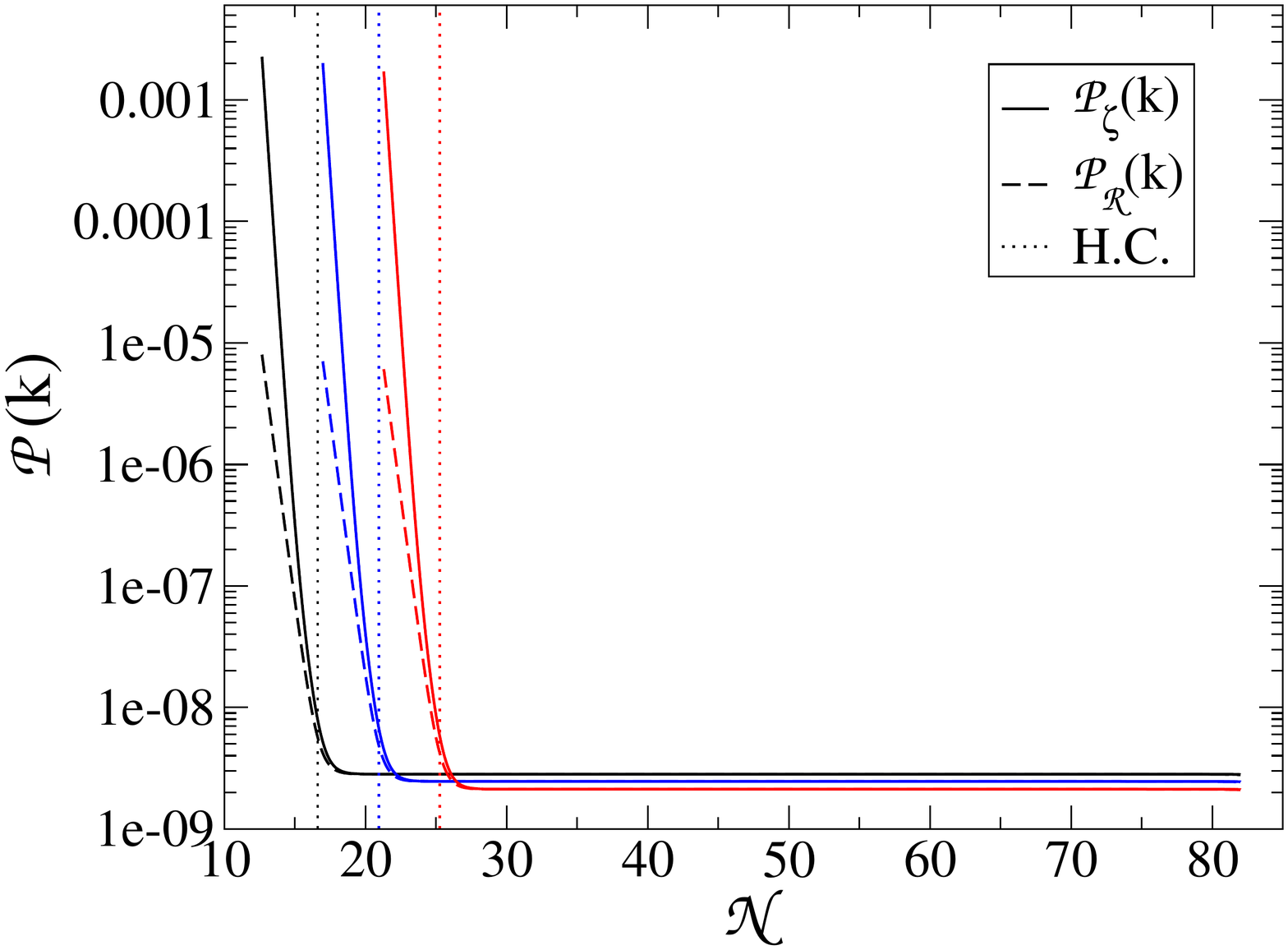}}}  \quad
      \subfigure
[\label{fig:1b} The evolution of $\dPnad$ is plotted against the number of efolds, $\N$. $\dPnad$ rapidly drops towards zero after horizon crossing.
We apply a cut off to the graph at $10^{-20}$, beneath which numerical noise dominates.]
{\scalebox{0.5}{\includegraphics[width=160mm]{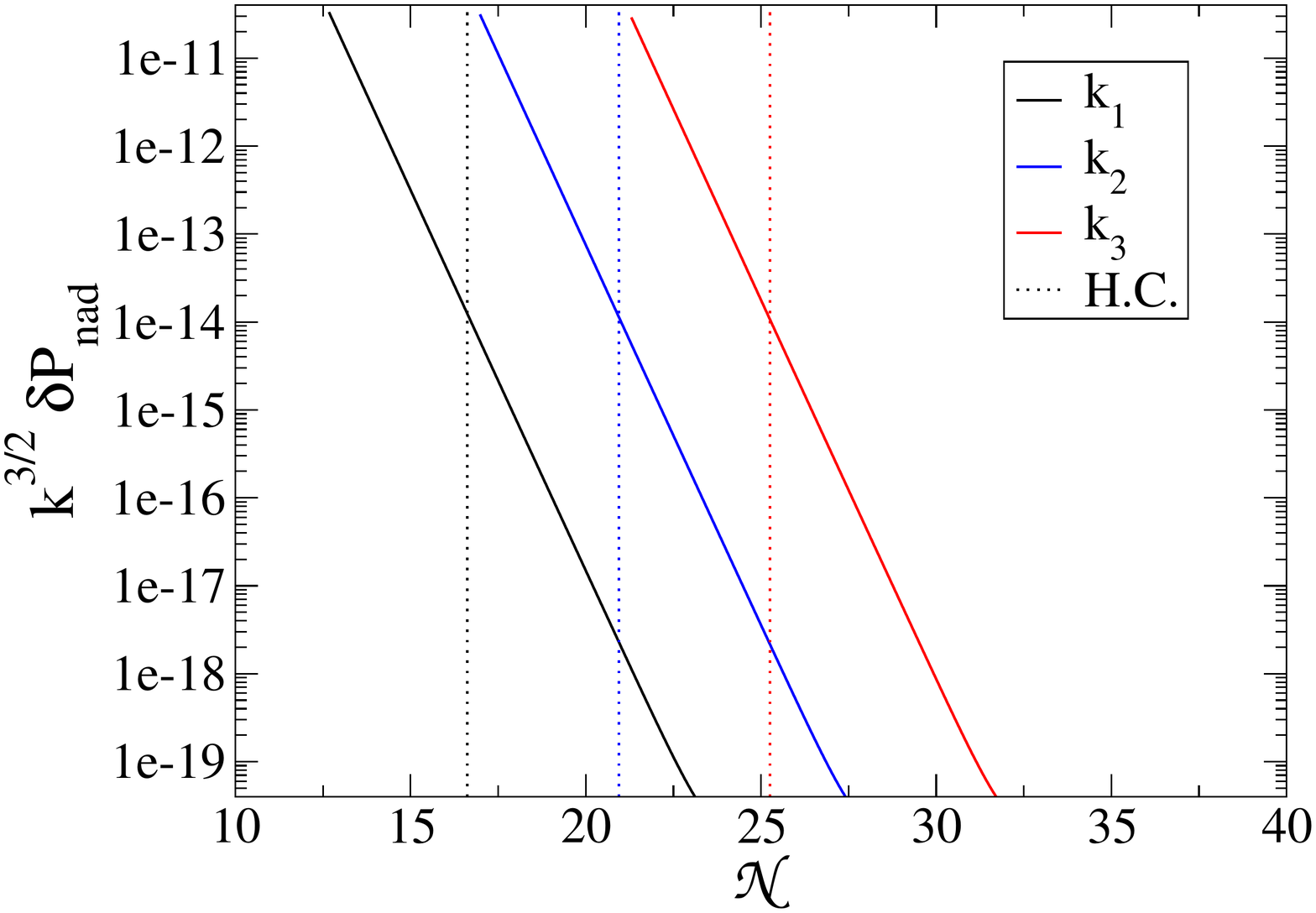}}} 
      }
    \caption{Both the graphs above are plotted for three different $k$ modes. The black line (left) is $k_1$, the blue line (middle) is $k_2$,
the WMAP pivot scale, and the red line (right) is $k_3$. }
\label{fig:evo}
  \end{center}
\end{figure}

Before discussing our results in detail it is worth considering the
evolution of the power spectra in general. The evolution, or indeed
conservation, of the curvature perturbations has been studied in
detail in the past, see Ref. \cite{Bardeen83, Lyth85,astro-ph/9511029,WMLL} and,
as expected, we find that some time shortly after horizon crossing
there is no longer any appreciable evolution in either the power
spectrum of $\zeta$, $\P_{\zeta}(k)$ or the power spectrum of $\R$,
$\P_{\R}(k)$, see Fig. \ref{fig:1a}. The values of these power
spectra converge very quickly onto the same conserved value. This
behaviour is also supported by the second graph in Fig. \ref{fig:1b}
which shows the non-adiabatic pressure perturbation, $\dPnad$. We
find, again as expected, that during and after horizon crossing the
size of $\dPnad$ drops sharply towards zero.  The non-adiabatic
pressure perturbation is directly related to the curvature
perturbation (on large scales), see Ref. \cite{astro-ph/9511029,WMLL} 
\be
{\zeta}' \propto \dPnad\, ,
\ee
and hence the rapid decrease in the non-adiabatic pressure
perturbation causes the curvature perturbations to settle onto a
conserved value.

However, it can also be seen from Fig. \ref{fig:1a} that there is
some evolution of the power spectra immediately after horizon
crossing.  This is well known in the literature where the phrase `soon
after horizon crossing' is commonly used to refer to the time at which
the power spectra have settled down.  In the sections that follow we
will be looking at this evolution in more detail and in particular
quantifying exactly how soon after horizon crossing the power spectra
reach the final value and how different this is to the horizon crossing values.

\subsection{How do $\zeta$ and $\R$ differ?}

The curvature perturbation on uniform density hypersurfaces, $\zeta$
and the comoving curvature perturbation, $\R$ are often used
interchangeably.  Although they have different definitions, see
\eq{defzeta} and \eq{defR}, it is well known that on large scales they
are equivalent, as can be seen from \eq{constraint}. This equivalence
is however only strictly true in the large scale limit, and on
smaller, finite scales this is not the case. In Fig. \ref{fig:1a} we
see that there is a difference between the two curvature perturbations
near to horizon crossing. Three different $k$-modes are plotted
throughout there evolution, from deep within the horizon through
horizon crossing (indicated by the dotted lines), until the end of
inflation.

\begin{figure}[ht!]
\includegraphics[width=80mm]{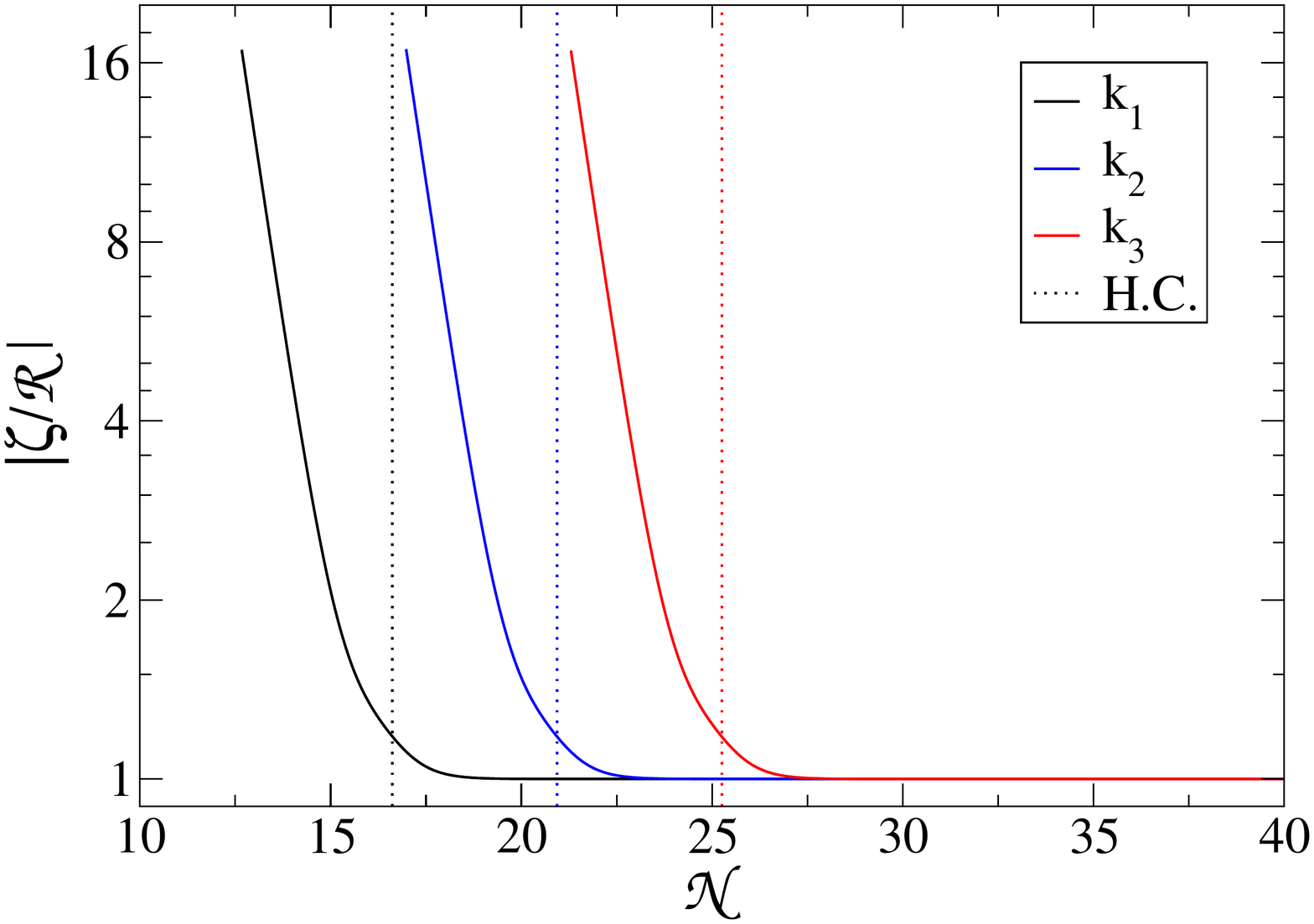} 
\caption{The ratio of $\zeta$ and $\R$ is plotted against the number of efolds, $\N$. Evolution is visible for a short period after horizon crossing,
during which time $\zeta$ is noticeably larger than $\R$.(Black line, left: $k_1$, Blue line, middle: $k_2$, Red line, right: $k_3$) }
\label{fig:zetaoverR}
\end{figure}

Fig. \ref{fig:zetaoverR} shows that $\zeta$ is as much as 20\%
larger than $\R$ at horizon crossing and remains significantly larger
for at least a couple of efolds. This highlights the importance of
making explicit the choice of curvature perturbation when carrying out
calculations close to horizon crossing. We also note that $\zeta$ and
$\R$ take slightly different amounts of time to settle down after
horizon exit, as detailed in Section \ref{sect:howquickly}.

\subsection{What is the error incurred by using horizon crossing quantities?}

\begin{figure}[ht!] 
  \begin{center}
    \mbox{
      \subfigure[\label{fig:pcR} $\P_{\zeta}(k)$]{\scalebox{0.5}{\includegraphics[width=160mm]{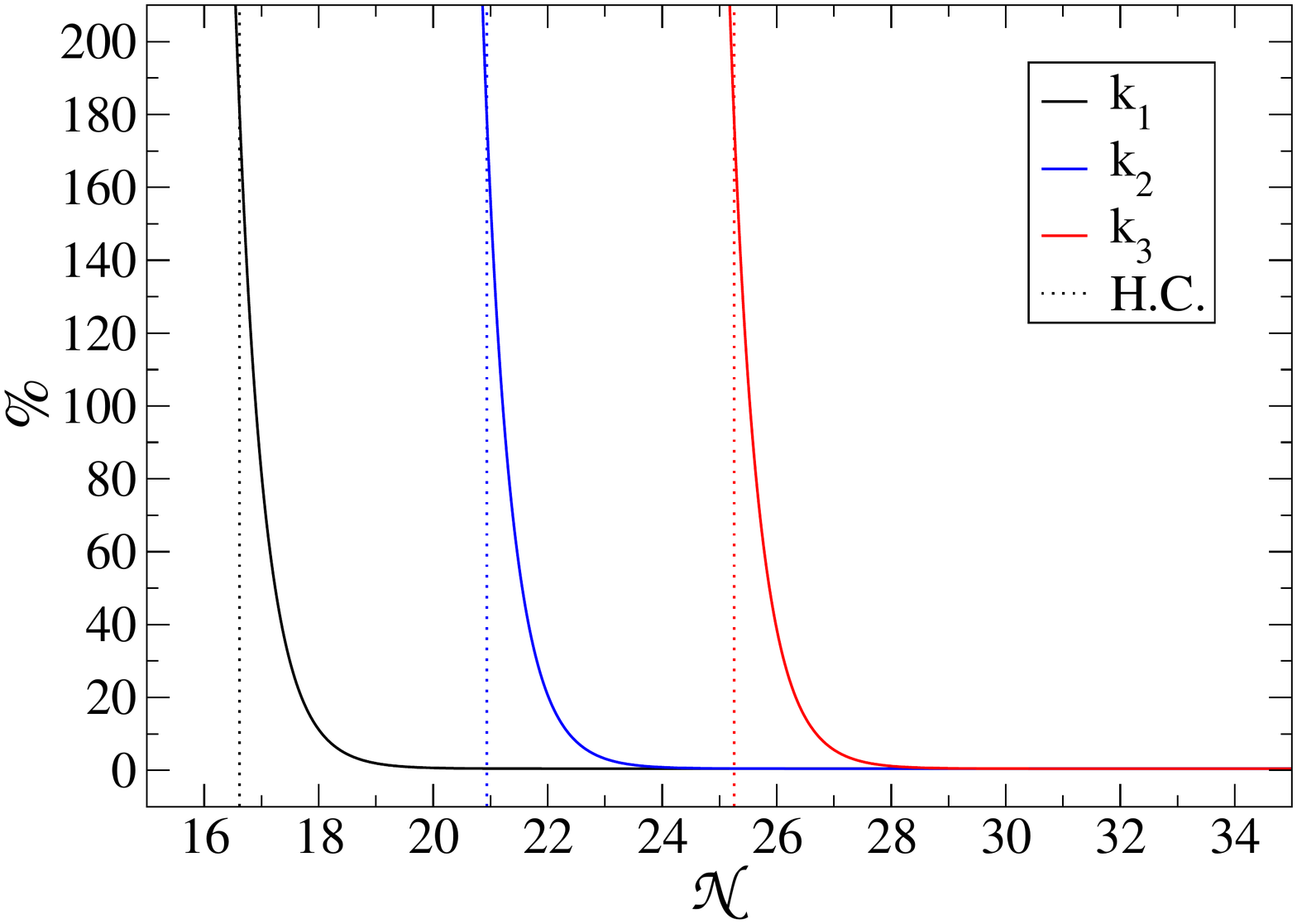}}}  \quad
      \subfigure[\label{fig:pczeta} $\P_{\R}(k)$]{\scalebox{0.5}{\includegraphics[width=160mm]{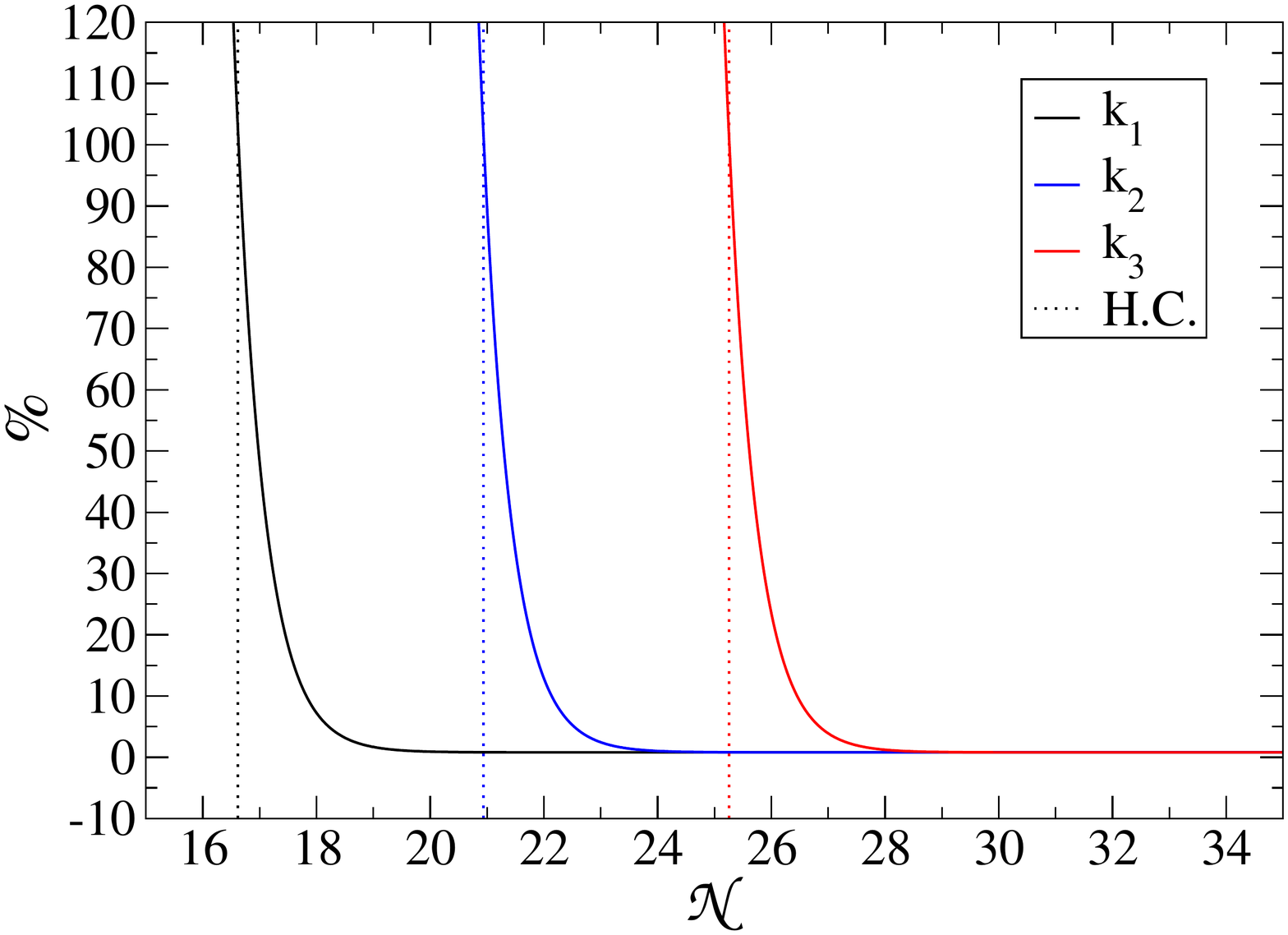}}} 
      }
    \caption{The percentage difference between $\P(k)$ evaluated, numerically, at the end of inflation and the $\P(k)$ obtained at each time step.
There is as much as 180\% difference in the values of $\P_{\zeta}(k)$ and 100\% difference in the values of $\P_{\R}(k)$.
Both the graphs above are plotted for three different $k$ modes. The black line (left) is $k_1$, the blue line (middle) is $k_2$,
the WMAP pivot scale, and the red line (right) is $k_3$.}
    \label{fig:pcchange}
  \end{center}
\end{figure}

As mentioned above it is well known that the values of the curvature
perturbation power spectra are not the same at horizon crossing as
they are at the end of inflation. However, it is not clear exactly how
much of an error would be incurred if one would use the horizon crossing
values instead of the correct values at the end of inflation.  In
Fig. \ref{fig:pcchange} we can see exactly how different the power
spectra are at horizon crossing compared to the values they take at
the end of inflation.  In Fig. \ref{fig:pcR} $\P_{\R}(k)$ is 100\%
larger at horizon crossing and in Fig. \ref{fig:pczeta}
$\P_{\zeta}(k)$ is 180\% larger.  
In fact if we use an analytic approximation that does take into account 
the pre horizon behaviour we can see the 100\% difference in the power 
spectra. Near horizon crossing the scalar field's wavefunction ($\psi$) is 
approximately proportional to $(1-ik\eta)e^{ik\eta}$. At horizon crossing $|k\eta| = 1$ and 
$|\psi|^2 \propto |1-ik\eta| = 2$.  A few e-folds later, $|k\eta| \approx 0$ and 
$|\psi^2| \propto |1| = 1$. This is a drop of 1. Hence the power spectra at 
horizon crossing is expected to be roughly 100\% larger than that at late times. 
The factor of two difference in $\P_{\R}(k)$, has previously been found analytically, 
for example Ref. \cite{Polarski:1995} 
This is, however, a large difference and might impact on calculations if 
$\P(k)_{k=aH}$ is used to approximate the value of the power spectra at 
the end of inflation.

\subsection{How quickly does the power spectra reach its final value at the end of inflation?}
\label{sect:howquickly}

\begin{figure}[ht!] 
  \begin{center}
    \mbox{
      \subfigure[$\P_{\zeta}(k)$]
{\scalebox{0.5}{\includegraphics[width=160mm]{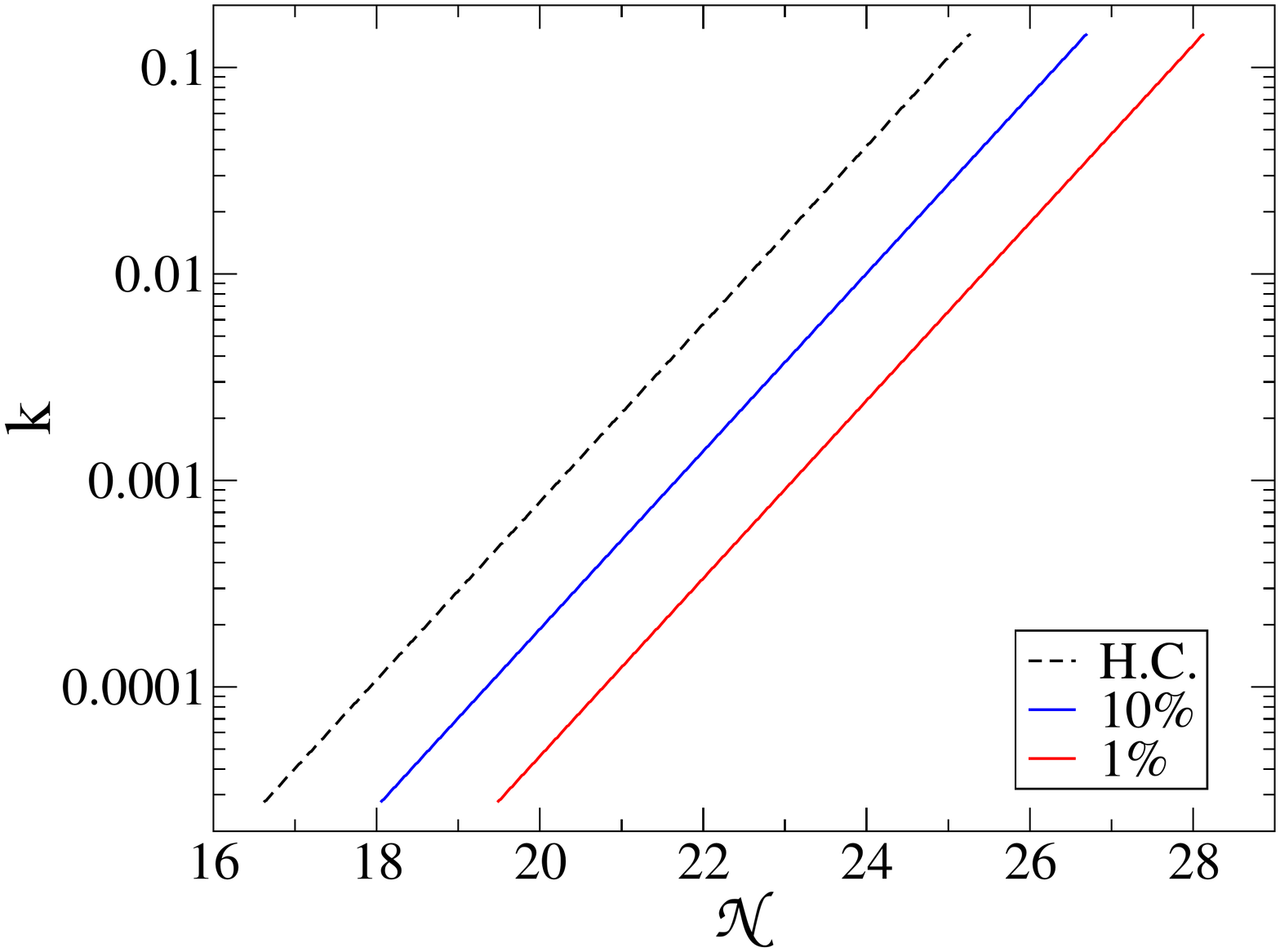}}} \quad
      \subfigure[ $\P_{\R}(k)$] 
{\scalebox{0.5}{\includegraphics[width=160mm]{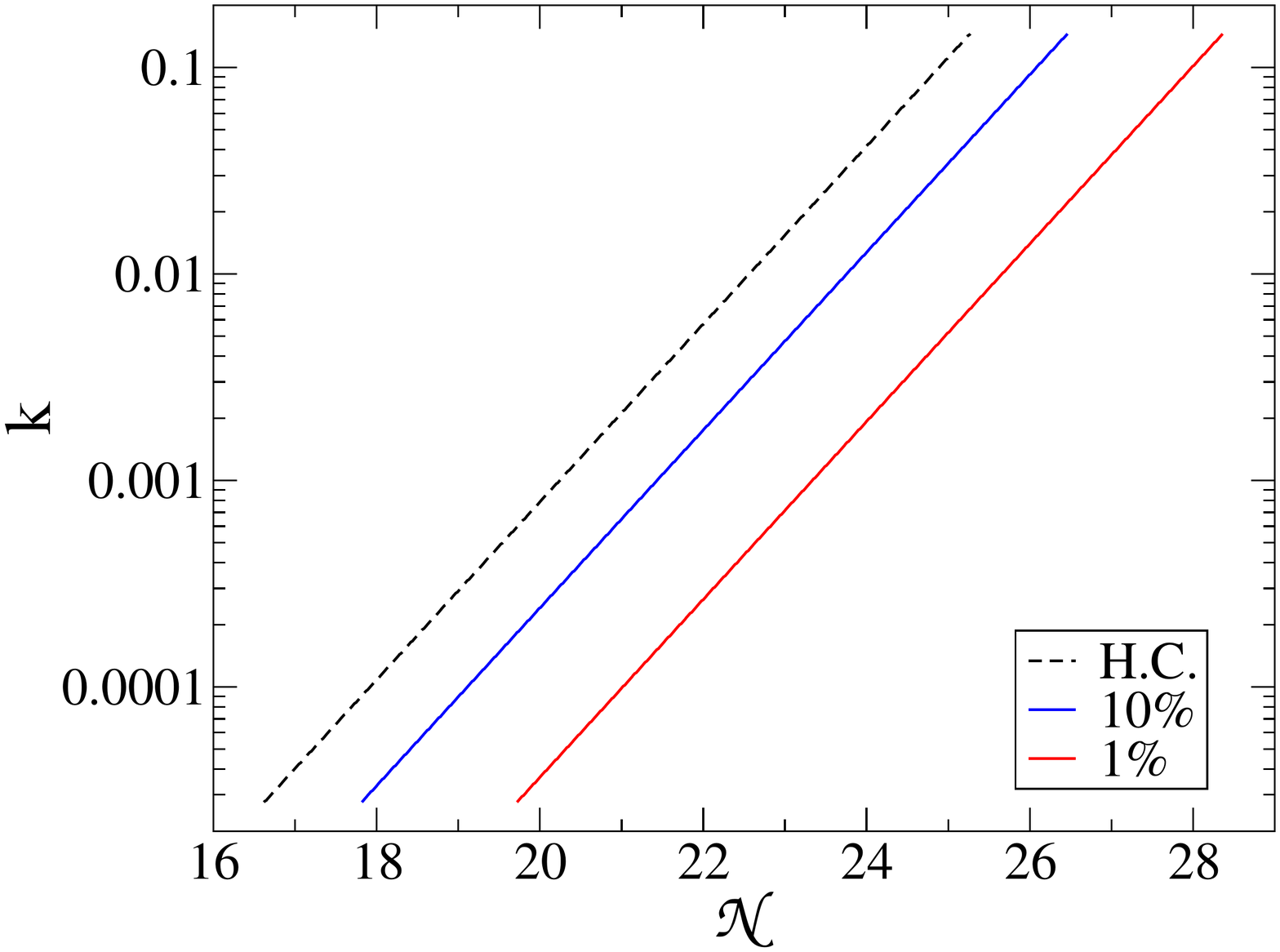}}}
      }
    \caption{These graphs show how many efolds one needs to wait for $\P (k)$ to be within 10\% and 1\% of $\P (k)$ at the end of inflation for a given value of $k$ (measured in $Mpc^{-1}$). 
Note that this is independent of $k$. Table \ref{table:1} gives the exact number of efolds for various potentials.}
    \label{fig:pclines}
  \end{center}
\end{figure}

The expressions ``soon after horizon crossing'' and ``a few efolds
after horizon crossing'' are often used in the literature to determine
when it is reasonable to evaluate the power spectra, so that one can
be confident that they have the same value as at the end of
inflation.  Now that we have established that the values at horizon
crossing can be as much as 180\% greater than those at the end of
inflation we can study exactly how long after horizon crossing we
must wait until the power spectra converge onto these final values.
Fig. \ref{fig:pclines} shows how many efolds one needs to wait for
the power spectra to be within 10\% and 1\% of the value they have at
the end of inflation.  It is clear from these graphs that this is
independent of $k$ and is in the region of ``a few efolds''.  Table
\ref{table:1} shows the number of efolds it takes to be within a
fixed percentage of the final value, and also shows data for the three
additional potentials we investigated. The choice of potential makes
very little difference to the result.  For every potential considered 
it took 1.40
-- 1.44 efolds for $\P_{\zeta}(k)$ to be within 10\% and 2.56 -- 2.86
to be within 1\%.  It took 1.14 -- 1.20 efolds for $\P_{\R}(k)$ to be
within 10\% and 2.30 -- 3.21 efolds to be within 1\%.  As we
highlighted in section \ref{sec:introduction} observational data will
soon be constraining observables to within a percent, so this is at least the
accuracy we would like to be able to evaluate quantities to.  Our
results show that to be within 1\% of the correct power spectra value
at the end of inflation, evaluating $\P_{\R}(k)$ approximately 3.2
efolds after horizon crossing and evaluating $\P_{\zeta}$ 2.9 efolds
after horizon crossing, would be sufficient, for all the potentials we
studied.

\begin{table} [ht!]
\renewcommand{\arraystretch}{1.3}
\begin{tabular}{@{} rrrrrcrrrr @{}} 
\hline
& \multicolumn{4}{c}{$\zeta$} & \phantom{abc} & \multicolumn{4}{c}{$\R$} \\
 & \quad 10\% \quad & \quad 5\% \quad & \quad 3\% \quad & \quad 1\% \quad && \quad 10\% \quad & \quad 5\% \quad & \quad 3\% \quad & \quad 1\% \quad \\
\hline
$\frac{1}{2}m^2\vp^2$ & 1.43 & 1.80 & 2.09 & 2.86 && 1.20 & 1.59 & 1.91 & 3.10 \\
$U_0 + \frac{1}{2}m^2\vp^2$ & 1.40 & 1.75 & 2.01 & 2.56 && 1.14 & 1.49 & 1.75 & 2.30 \\
$\frac{1}{4} \lambda \vp^4$ & 1.44 & 1.80 & 2.08 & 2.77 && 1.18 & 1.55 & 1.84 & 2.59 \\
$\sigma \vp^{2/3}$ & 1.41 & 1.77 & 2.05 & 2.74 && 1.20 & 1.61 & 1.98 & 3.21 \\
\hline
\end{tabular}
\caption{The values in this table represent how many efolds after horizon crossing it takes for the power spectrum 
to be within a fixed percentage of the power spectrum at the end of inflation.}
\label{table:1}
\end{table}

\subsection{Highlighting the magnitude of the error made in using the incorrect analytic expression}
\label{sect:highlight}

\begin{figure}[ht!]
  \begin{center}
    \mbox{
      \subfigure[The lines in this graph are shown for the mode $k_2$ which is the WMAP pivot scale.]
{\scalebox{0.5}{\includegraphics[width=160mm]{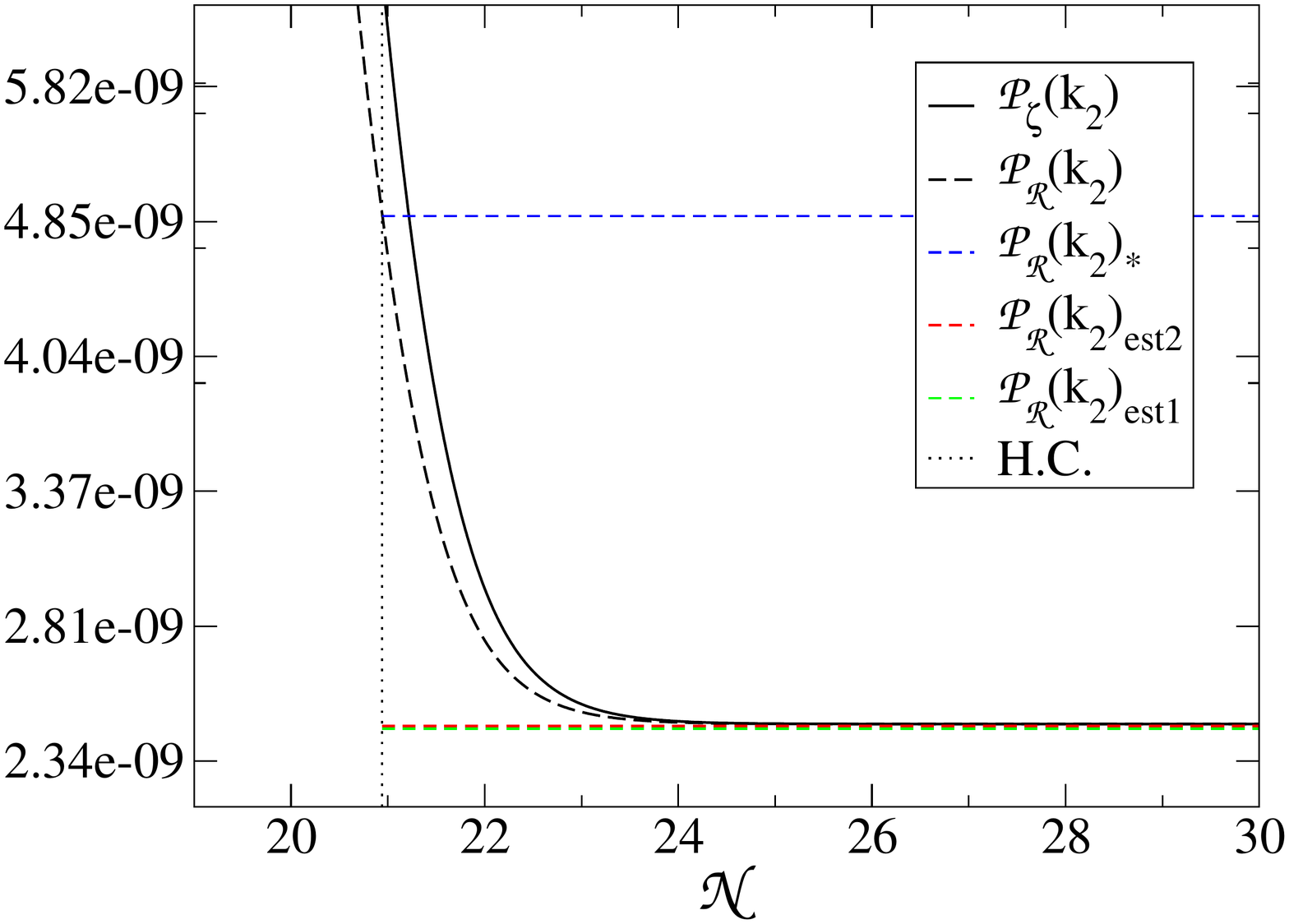}}} \quad
      \subfigure[\label{comparepic2}The lines in this graph are shown for the mode $k_2$ which is the WMAP pivot scale (left hand lines) and $k_3$ (right hand lines).] 
{\scalebox{0.5}{\includegraphics[width=160mm]{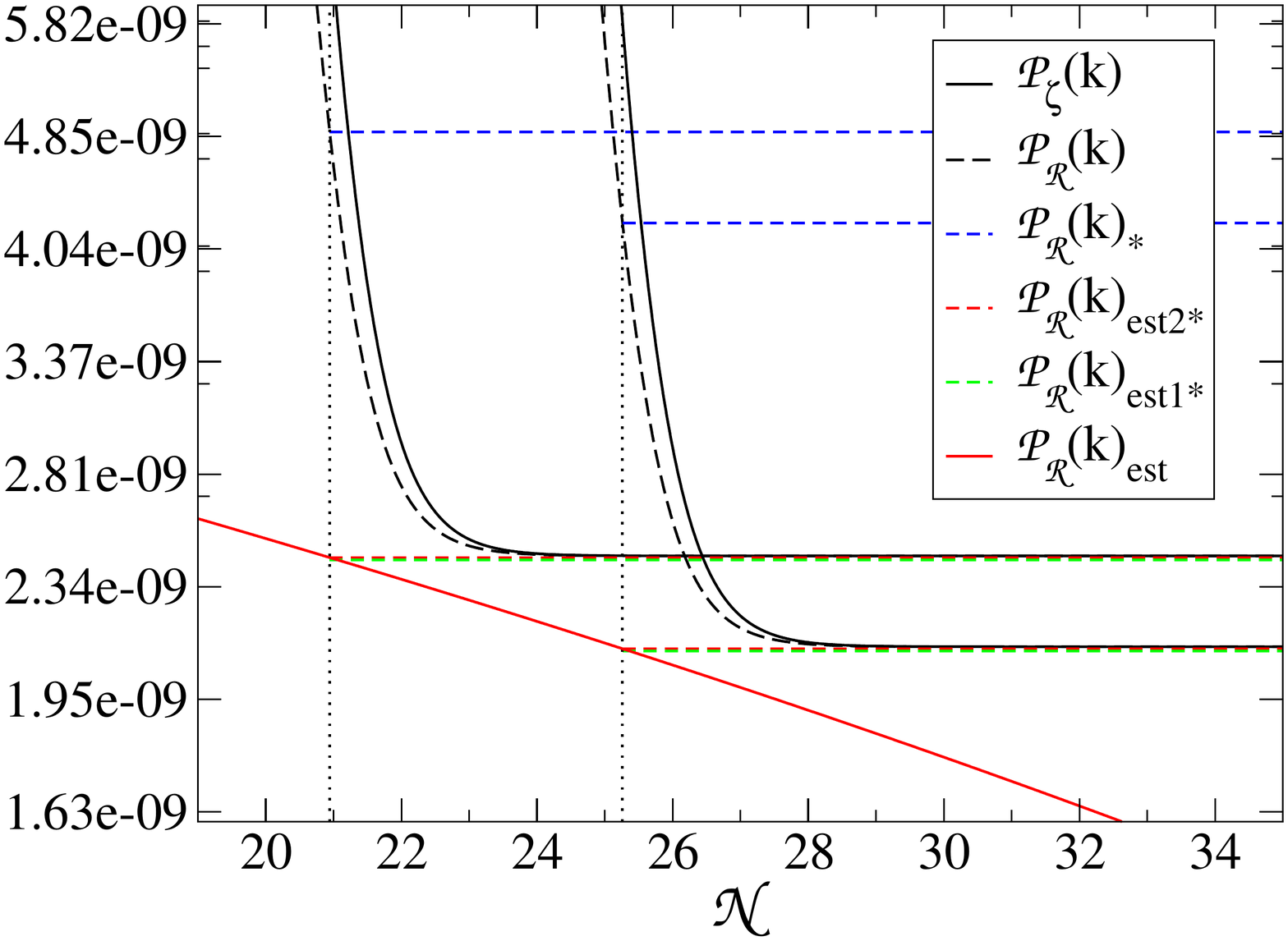}}}
      }
    \caption{The evolution of the two types of curvature perturbation, $\P_{\R}(k)$ and $\P_{\zeta}(k)$ are plotted against the number of efolds, $\N$.
These numerical solutions are compared to the correct analytic solution with and without slow roll corrections (\eq{PRest2} and \eq{PRest1}
and the na\"ive approximation of evaluating the power spectra at horizon crossing.}
    \label{fig:compare}
  \end{center}
\end{figure}

In most analytic calculations an expression for the power spectrum is
derived using the $k \to 0$ limit but is evaluated using quantities at
horizon crossing, as given in \eq{PRest2}. As we mentioned in
section \ref{sec:introduction}, it is often not made explicitly clear
in the literature when the mixing of late time solution and horizon
crossing values is being used, even when this is well understood by
the authors.  In Fig. \ref{fig:compare} we compare both this correct
analytic solution and the na\"ive calculation of the power spectrum at
horizon crossing performed using horizon crossing values with the
numerical solutions. In Fig. \ref{fig:compare}, as expected, the
correct analytic solution gives a very good estimate to the full
numerical solution. Even the analytic solution without the slow roll
correction, given in \eq{PRest1}, is a very good estimate to the
full solution, in fact the error in not including the slow roll
corrections for the WMAP scale is only a slight underestimate of
0.38\%. However, when we compare the numerical solution to
$\P_{\R}(k)$ evaluated at horizon crossing we find, as shown in Fig.
\ref{fig:pcR} that there is a 100\% error in our answer. As we have
shown earlier in Fig. \ref{fig:pczeta}, if we evaluated the
$\P_{\zeta}(k)$ at horizon crossing the error would be even higher at
180\%, this difference in the two power spectra can be seen clearly in
Fig. \ref{fig:compare}. This again highlights the importance of
establishing that the correct analytic expression is used in
calculations, and when different calculations and results are compared.

Another possible source of error would be to use the analytic
expression given in \eq{PRest2} above, but not to evaluate it at
horizon crossing. If one were to evaluate this expression `some efolds
after horizon crossing', one would underestimate the amplitude of the
power spectrum. This corresponds to following the red line in Fig.
\ref{comparepic2}. For example, evaluating \eq{PRest2} 4 efolds after
horizon crossing would incur an error of 15\%. Furthermore, evaluating
the power spectrum at later and later times will increase the error.
Lastly, it is worth noting that although the analytic and numerical expressions
agree in the large scale limit, they do not agree with each other shortly after horizon crossing.
As shown in the section above one must wait at least 3.2 efolds for these two values to agree. 
This is particularly important if there is a second phase of evolution caused for example by a second scalar field which   
starts to dominate during these three efolds, see for instance Ref. \cite{Kinney:2005}.

\subsection{The Spectral Index}
\label{sect:spectral}

\begin{figure}[ht!]
\includegraphics[width=80mm]{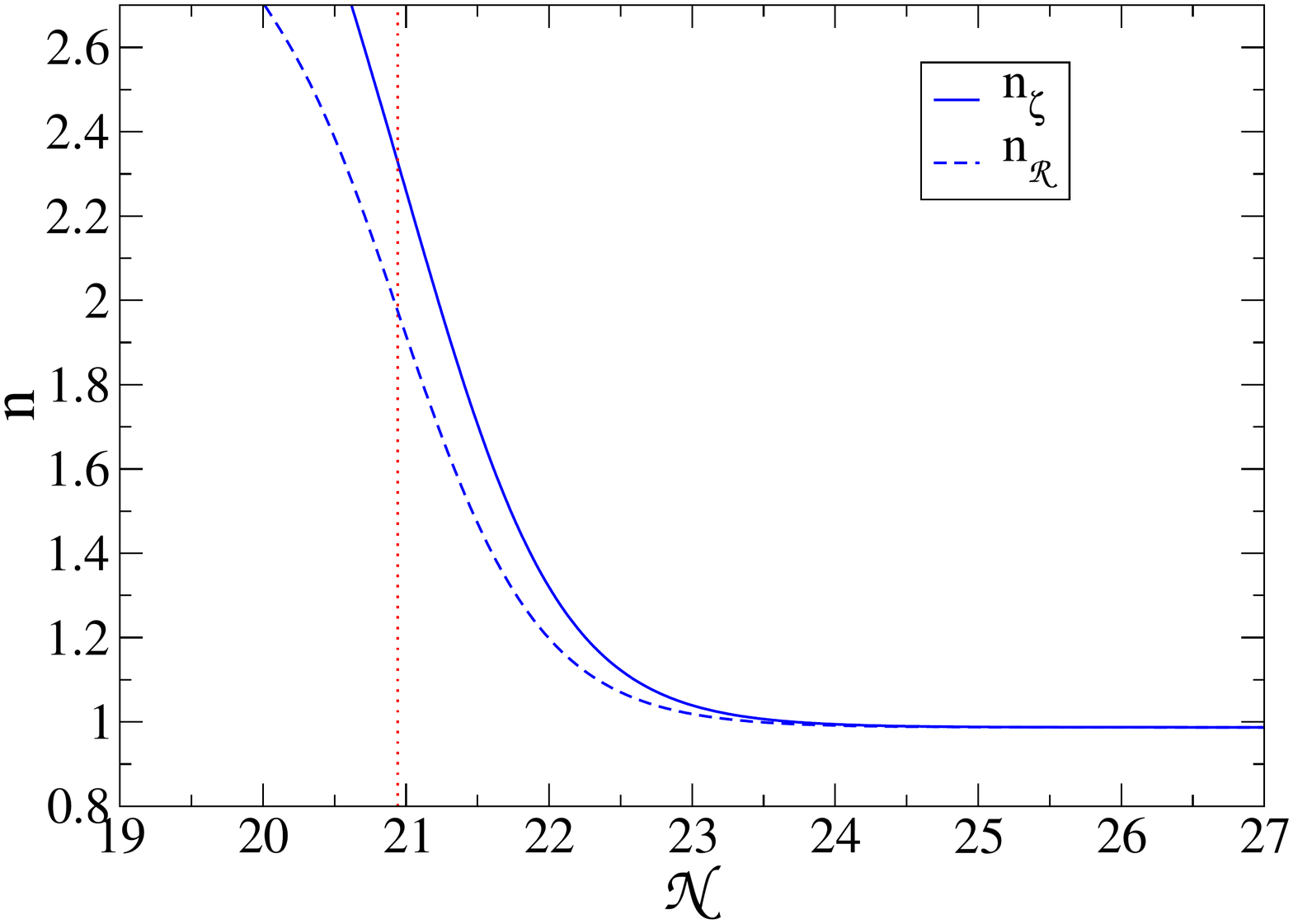} 
\caption{The evolution of the spectral index for the two curvature perturbations, 
$n_{\R}$ and $n_{\zeta}$, are plotted against the number of efolds, $\N$. 
The lines in this graph are shown for the mode $k_2$ which is the WMAP pivot scale.}
\label{fig:spectral}
\end{figure}

Using observations we can gain information about the power spectrum of curvature perturbations, which in turn allows us to constrain cosmological theories.
Many observational results are given in terms of a few observables which can then be compared directly with predictions given by theories, one such 
observable is the spectral index. The spectral index describes the scale dependence of the power spectrum of the curvature perturbation and is defined by \cite{LLBook}
\be
n_{\zeta} - 1 \equiv \frac{d\rm{ln}\P_{\zeta}(k)}{d\rm{ln}k} \, , \qquad  n_{\R} - 1 \equiv \frac{d\rm{ln}\P_{\R}(k)}{d\rm{ln}k} \, .
\ee

As an example of how the results presented in this paper will impact on particular observables we consider the spectral index in more detail, we present
results here using chaotic inflation. 
Figure \ref{fig:spectral} shows, as expected, that just like the curvature perturbation the spectral index continues to evolve for a few efolds after
the mode has crossed outside the horizon. Evaluating the spectral index naively at horizon crossing gives a results of $n_{\R} \approx 2$ and 
$n_{\zeta} \approx 2.3$, which is an error in both cases of more than 100\%. We find that in order for $n_{\zeta}$ and $n_{\R}$ to be within 1\% of their 
values at the end of inflation, we should evaluate them at least 2.91 and 2.66 efolds after horizon crossing, respectively. These values are 
similar but slightly less than the number of efolds it takes for the power spectrum to be within 1\% of it's final value, see Table 1.

\section{Conclusions}
\label{sec:conclusions}

In this paper we have quantified the evolution of the curvature
perturbations after inflation and highlighted possible errors which
can occur. As we are entering an era where we can hope to constrain
cosmological parameters to within a percent using the observational
data from e.g.~Planck, it is of particular importance that these
errors are both minimised and quantified.  

After presenting the evolution equations for the scalar field, we gave
the expressions for the curvature perturbations we consider. We then
outlined the numerical methods used to solve these equations and gave
details of the various models and the initial conditions used. In
presenting our results, we found that despite $\zeta$ and $\R$ being
equivalent very far outside the horizon, the difference between $|\R|$
and $|\zeta|$ at horizon crossing can be as much as 20\%.  We also
found the error in evaluating the power spectra numerically at horizon crossing
instead of either using the correct analytic expression or the full
numerical solution at late times can be as much as 180\% for $\P_{\zeta}(k)$ and
100\% for $\P_{\R}$.  Lastly we showed that if one wanted to evaluate
the power spectra without the use of the analytic expression
\eq{PRest2}, one would need to wait at least 3.2 efolds to ensure the
answer for $\P_{\R}(k)$ is correct to within 1\% of the value at the
end of inflation, and one would need to wait at least 2.9 efolds to
ensure the answer for $P_{\zeta}$ is correct to within 1\%.  
There was no significant difference to these results when we considered the
three additional single field models presented at the end of Section
\ref{sec:numerical}. 

In this paper we point out that there is a difference between analytic 
and numerical expressions close to the horizon.
This is due to the fact that the long wavelength approximation is not
valid immediately on horizon crossing. That is, there are terms
in the analytical solution involving gradients that are non-negligible 
on horizon crossing. These become negligible rapidly, and after
around 3 e-folds the exact numerical solution agrees with the 
long-wavelength analytical approximations.
Furthermore, from Fig.~\ref{comparepic2}, we can see that these corrections are not
due to the break in slow roll (since the two analytical estimates are roughly the same),
but that this difference really arises due to the invalidity of the long
wavelength approximation on horizon crossing.

The numerical results 
are the instantaneous values of the power spectrum and spectral index 
at horizon crossing, not the expected late time values. These instantaneous 
results, while not of observational significance, are useful in many ways, 
including as initial conditions for other analytical or numerical schemes 
which operate purely outside the horizon. Firstly, if we are interested 
in the late time values we should not take the `na\"{\i}ve' numerical approach 
of evaluating these at horizon crossing. Unlike the case when using the 
analytic expressions these results will not be close to the correct answer. 
Secondly, if we are interested in the instantaneous values at or close to 
horizon crossing, for instance when developing codes which rely on this 
information, the normal analytic expressions will not give the correct 
answers as they are no longer valid and you need to use numerical methods. 

In this paper we have only studied single field
inflation models, in which the non-adiabatic pressure decays rapidly,
see Fig.~\ref{fig:1b}. This is no longer the case in multi-field
systems, as a recent work has detailed, Ref. \cite{CH2011}. 
It will be interesting to repeat our analysis for more complicated models 
where superhorizon evolution of the curvature
perturbation is expected. This would include, for instance multi-field 
inflation or ``ultra slow roll inflation'', see for instance Ref. 
\cite{Kinney:2005, Leach:2001}. Since in this paper we focus on 
the single field case, we postpone a study of these cases for future work.

We have chosen the end of inflation as a natural end point of our
calculations. After the end of inflation the inflaton is assumed to
decay into the standard matter fields during reheating, the detailed
mechanism of which is as yet unclear. Also, after reheating we have a
multi-fluid system, and hence automatically $\dPnad\neq 0$, see Ref.
\cite{BCM2011}, which means that the curvature perturbations are no longer
conserved (on any scale). Consequently the values at the end of
reheating might no longer be the same as those at the end of
inflation.

In conclusion, we have highlighted the fact that the different
curvature perturbations do evolve differently immediately after
horizon exit. Confusion of the different curvature perturbations can
introduce additional errors when comparing theoretical results with
observations, which can easily be avoided.

\section*{Acknowledgements}

We would like to thank David Seery and Will Kinney for useful comments and discussions.
EN is funded by a STFC studentship. AJC is funded by the Sir Norman
Lockyer Fellowship of the Royal Astronomical Society, IH is funded by
the STFC under Grant ST/G002150/1, and KAM is supported in part by the
STFC under Grants ST/G002150/1 and ST/H002855/1.

{}


\begin{thebibliography}{}

\bibitem{Kolbandturner}
 E.~W.~Kolb and M.~S.~Turner,
Front.\ Phys.\ \ {\bf 69}, 1  (1990).

\bibitem{LLBook}
A.~R.~Liddle and D.~H.~Lyth,
\emph{Cosmological inflation and large-scale structure}, CUP,
Cambridge, UK (2000).

\bibitem{WMAP1}
  E.~Komatsu {\it et al.} [WMAP Collaboration],
  Astrophys.\ J.\ Suppl.\ \ {\bf 148}, 119  (2003)
  [astro-ph/0302223].

\bibitem{WMAP7}
  E.~Komatsu {\it et al.} [WMAP Collaboration],
  Astrophys.\ J.\ Suppl.\ \ {\bf 192}, 18  (2011)
  [arXiv:1001.4538 [astro-ph.CO]].

\bibitem{Bardeen83}
J.~M.~Bardeen, P.~J.~Steinhardt and M.~S.~Turner,
Phys.\ Rev.\ D {\bf 28}, 679 (1983).


\bibitem{Lyth85}
D.~H.~Lyth,
Phys.\ Rev.\ D {\bf 31}, 1792 (1985).

\bibitem{WMLL}
D.~Wands, K.~A.~Malik, D.~H.~Lyth and A.~R.~Liddle,
Phys.\ Rev.\ D {\bf 62}, 043527 (2000)
[arXiv:astro-ph/0003278].

\bibitem{Copeland:1993jj} 
  E.~J.~Copeland, E.~W.~Kolb, A.~R.~Liddle and J.~E.~Lidsey,
Phys.\ Rev.\ D\ {\bf 48}, 2529  (1993)
[hep-ph/9303288].
��

\bibitem{Mollerach}
  S.~Mollerach,
  Phys.\ Rev.\  D {\bf 42}, 313 (1990).


\bibitem{Stewart:1993}  
  E.~D.~Stewart and D.~H.~Lyth,
��Phys.\ Lett.\ B\ {\bf 302}, 171  (1993)
��[gr-qc/9302019].
��


\bibitem{Grivell} 
  I.~J.~Grivell and A.~R.~Liddle,
Phys.\ Rev.\ D\ {\bf 54}, 7191  (1996)
[astro-ph/9607096].
��

\bibitem{Huang:2000bh} 
  D.~H.~Huang, W.~B.~Lin and X.~M.~Zhang,
Phys.\ Rev.\ D\ {\bf 62}, 087302  (2000)
[hep-ph/0007064].
��

\bibitem{Leach:2001zf}
  S.~Leach, M.~Sasaki, D.~Wands and A.~R.~Liddle,
Phys.\ Rev.\ D\ {\bf 64}, 023512  (2001)
[astro-ph/0101406].
��

\bibitem{Leach:2001} 
  S.~M.~Leach and A.~R.~Liddle,
  Phys.\ Rev.\ D\ {\bf 63}, 043508  (2001)
  [astro-ph/0010082].


\bibitem{Stewart:2002} 
  E.~D.~Stewart,
��Phys.\ Rev.\ D\ {\bf 65}, 103508  (2002)
��[astro-ph/0110322].
��


\bibitem{J.Lidsey} 
  J.~E.~Lidsey, A.~R.~Liddle, E.~W.~Kolb, E.~J.~Copeland, T.~Barreiro and M.~Abney,
  Rev.\ Mod.\ Phys.\ \ {\bf 69}, 373  (1997)
  [astro-ph/9508078].

\bibitem{hep-ph/9807278} 
  D.~H.~Lyth and A.~Riotto,
  Phys.\ Rept.\ \ {\bf 314}, 1  (1999)
  [hep-ph/9807278].

\bibitem{MW2008}
  K.~A.~Malik and D.~Wands,
  Phys.\ Rept.\  {\bf 475}, 1 (2009)
  [arXiv:0809.4944 [astro-ph]].

\bibitem{Polarski:1995} 
  D.~Polarski and A.~A.~Starobinsky,
  Class.\ Quant.\ Grav.\ \ {\bf 13}, 377  (1996)
  [gr-qc/9504030].

\bibitem{Lyth:2006} 
  D.~H.~Lyth and D.~Seery,
  Phys.\ Lett.\ B\ {\bf 662}, 309  (2008)
  [astro-ph/0607647].



\bibitem{Sasaki1986}
M.~Sasaki,
Prog.\ Theor.\ Phys.\  {\bf 76}, 1036 (1986).

\bibitem{Mukhanov88}
V.~F.~Mukhanov,
Sov.\ Phys.\ JETP {\bf 67}, 1297 (1988)
[Zh.\ Eksp.\ Teor.\ Fiz.\  {\bf 94N7}, 1 (1988)].



\bibitem{camb}
A.~Lewis and A.~Challinor
{\tt{http://camb.info/}}



\bibitem{KS}
H.~Kodama and M.~Sasaki,
Prog.\ Theor.\ Phys.\ Suppl.\  {\bf 78}, 1 (1984).


\bibitem{nonad}
  A.~J.~Christopherson and K.~A.~Malik,
  Phys.\ Lett.\  B {\bf 675} (2009) 159
  [arXiv:0809.3518 [astro-ph]].

\bibitem{Starobinsky:1982ee}
  A.~A.~Starobinsky,
  Phys.\ Lett.\ B {\bf 117}, 175 (1982).


\bibitem{Starobinsky:1986fx}
  A.~A.~Starobinsky,
  JETP Lett.\  {\bf 42}, 152 (1985)
  [Pisma Zh.\ Eksp.\ Teor.\ Fiz.\  {\bf 42}, 124 (1985)].

\bibitem{Sasaki:1995aw}
M.~Sasaki and E.~D.~Stewart,
Prog.\ Theor.\ Phys.\  {\bf 95}, 71 (1996)
[arXiv:astro-ph/9507001].



\bibitem{LMS}
D.~H.~Lyth, K.~A.~Malik and M.~Sasaki,
JCAP {\bf 0505}, 004 (2005)
[arXiv:astro-ph/0411220].

\bibitem{Burrage:2011} 
  C.~Burrage, R.~H.~Ribeiro and D.~Seery,
  JCAP\ {\bf 1107}, 032  (2011)
  [arXiv:1103.4126 [astro-ph.CO]].

\bibitem{Kinney:2005} 
  W.~H.~Kinney,
  Phys.\ Rev.\ D\ {\bf 72}, 023515  (2005)
  [gr-qc/0503017].

\bibitem{Byrnes:2009} 
  C.~T.~Byrnes, S.~Nurmi, G.~Tasinato and D.~Wands,
  JCAP\ {\bf 1002}, 034  (2010)
  [arXiv:0911.2780 [astro-ph.CO]].


\bibitem{Salopek:1988qh}
D.~S.~Salopek, J.~R.~Bond and J.~M.~Bardeen,
Phys.\ Rev.\ D {\bf 40}, 1753 (1989).


\bibitem{Ianthesis}
  I.~Huston,
  \emph{``Constraining Inflationary Scenarios with Braneworld Models and Second Order Cosmological Perturbations,''}
  arXiv:1006.5321 [astro-ph.CO].

\bibitem{Huston:2009ac}
  I.~Huston and K.~A.~Malik,
  JCAP {\bf 0909}, 019 (2009)
  [arXiv:0907.2917 [astro-ph.CO]].


\bibitem{pyflation} 
  I.~Huston and K.~A.~Malik,
  JCAP\ {\bf 1110}, 029  (2011)
  [arXiv:1103.0912 [astro-ph.CO]].


\bibitem{astro-ph/9511029} 
  J.~Garcia-Bellido and D.~Wands,
  Phys.\ Rev.\ D\ {\bf 53}, 5437  (1996)
  [astro-ph/9511029].


\bibitem{CH2011} 
  I.~Huston and A.~J.~Christopherson,
  Phys.\ Rev.\ D {\bf 85} (2012) 063507
  [arXiv:1111.6919 [astro-ph.CO]].


\bibitem{BCM2011}
  I.~A.~Brown, A.~J.~Christopherson and K.~A.~Malik,
  Mon.\ Not.\ Roy.\ Astron.\ Soc.\  {\bf 423} (2012) 1411
  [arXiv:1108.0639 [astro-ph.CO]].






\end{thebibliography}
\end{document}